\documentclass[reprint,showpacs,preprintnumbers,amsmath,amssymb,nofootinbib,groupedaddress,superscriptaddress,showkeys,twocolumn]{revtex4-1}

\usepackage{graphicx}
\usepackage{dcolumn}
\usepackage{epsfig}
\usepackage[dvips]{color}
\usepackage{bm}
\usepackage{fancybox}
\usepackage{mathrsfs}
\usepackage{booktabs}
\usepackage{framed}
\definecolor{shadecolor}{gray}{0.90}
\usepackage{amsmath}	

\allowdisplaybreaks[4]

\begin{document}

\title{Mind the gap on Icecube: 
Cosmic neutrino spectrum and muon anomalous magnetic moment
in the gauged $L_{\mu} - L_{\tau}$ model}

\author{Takeshi Araki}
\email{araki@krishna.th.phy.saitama-u.ac.jp}
\affiliation{%
Department of physics, Saitama University,
\\
Shimo-Okubo 255, 
338-8570 Saitama Sakura-ku, 
Japan
}

\author{Fumihiro Kaneko}
\email{kaneko@krishna.th.phy.saitama-u.ac.jp}
\affiliation{%
Department of physics, Saitama University,
\\
Shimo-Okubo 255, 
338-8570 Saitama Sakura-ku, 
Japan
}

\author{Yasufumi Konishi}
\email{konishi@krishna.th.phy.saitama-u.ac.jp}
\affiliation{%
Department of physics, Saitama University,
\\
Shimo-Okubo 255, 
338-8570 Saitama Sakura-ku, 
Japan
}

\author{Toshihiko Ota}
\email{toshi@mail.saitama-u.ac.jp}
\affiliation{%
Department of physics, Saitama University,
\\
Shimo-Okubo 255, 
338-8570 Saitama Sakura-ku, 
Japan
}

\author{Joe Sato}
\email{joe@phy.saitama-u.ac.jp}
\affiliation{%
Department of physics, Saitama University,
\\
Shimo-Okubo 255, 
338-8570 Saitama Sakura-ku, 
Japan
}

\author{Takashi Shimomura}
\email{takashi@krishna.th.phy.saitama-u.ac.jp}
\affiliation{%
Department of physics, Saitama University,
\\
Shimo-Okubo 255, 
338-8570 Saitama Sakura-ku, 
Japan
}

\date{\today}

\pacs{
13.15.+g, 
14.60.Ef, 
95.55.Vj, 
98.70.Sa,
}

\preprint{\bf STUPP-14-219}


\begin{abstract}
The energy spectrum of cosmic neutrinos, which was recently 
reported by the {\sf IceCube} collaboration,
shows a gap between 400 TeV and 1 PeV.
An unknown neutrino interaction mediated by a field 
with a mass of the MeV scale is one of the possible 
solutions to this gap.
We examine if the leptonic gauge interaction 
$L_{\mu} - L_{\tau}$ can simultaneously explain 
the two phenomena in the lepton sector:
the gap in the cosmic neutrino spectrum 
and the unsettled disagreement 
in muon anomalous magnetic moment.
We illustrate that there remains the regions in the model parameter space, 
which account for both the problems.
Our results also provide a hint for the distance to 
the source of the high-energy cosmic neutrinos.
\end{abstract}

\maketitle

{\it Introduction --} Following the observations of the celebrated two
events~\cite{Aartsen:2013bka,Aartsen:2013jdh}, {\sf IceCube} has
accumulated 37 high-energy neutrino events which are 
significantly greater than the expected number of background events
originate from atmospheric neutrinos and cosmic 
ray muons~\cite{Aartsen:2014gkd}.
These events start telling us an energy spectrum of cosmic neutrinos 
at the uncharted high-energy regions.  The spectrum is consistent 
with the Waxman-Bahcall bound~\cite{Waxman:1998yy,Bahcall:1999yr} 
estimated from the high-energy cosmic-ray observations.  
An interesting and unexpected feature of the {\sf IceCube} spectrum 
is that there is a gap in the energy range between 400 TeV and 1 PeV. 
Although the existence of the gap has not been statistically 
established yet, some attempts to explain the gap have been
examined~\cite{Ng:2014pca,Ioka:2014kca,Ibe:2014pja,Blum:2014ewa}.  
An attractive candidate of the explanation is an attenuation process 
driven by an unknown interaction between the high-energy cosmic neutrino 
and the Cosmic neutrino Background (C$\nu$B).  This type of interaction,
{\it the secret neutrino interaction}~\cite{Bilenky:1999dn}, has been
discussed especially in the context of cosmology and astrophysics
\cite{Kolb:1987qy,Keranen:1997gz,Goldberg:2005yw,Baker:2006gm,Hooper:2007jr,Dolgov:1995hc}.
If the gap in the {\sf IceCube} spectrum suggests the novel leptonic
force, what can we expect as the other phenomenological consequences?
In this letter, we introduce the gauged $U(1)$ leptonic interaction
associated with {\it the muon number minus the tau number} 
$L_{\mu} - L_{\tau}$, which is anomaly-free within the Standard Model 
(SM) particle contents~\cite{Foot:1990mn,He:1990pn} and 
can naturally explain the large atmospheric mixing
\cite{Choubey:2004hn,Ota:2006xr,Heeck:2010pg,Heeck:2011wj}.
We examine if this new interaction can explain not only the gap 
in the cosmic neutrino spectrum, but also the long-standing 
inconsistency between experiments and theory
in the muon anomalous magnetic moment ($g_{\mu} -2$), whose statistical significance is about 3 $\sigma$~\cite{Hagiwara:2011af}.

On the one hand the cosmological and the experimental bounds
 to the secret neutrino interactions are not strict~\cite{Lessa:2007up},
but on the other hand those with large couplings are difficult to be
motivated from the theoretical point of view.  Because of the $SU(2)$
symmetry in the SM, the introduction of a secret
neutrino interaction, in general, results in providing an interaction to
the corresponding charged lepton with the size of the same order as the
neutrino interaction.  In Refs.~\cite{Ibe:2014pja,Blum:2014ewa}, the
authors brought an $SU(2)$ violation in their leptonic interaction to
circumvent the possible problems caused by the charged lepton sector of
their leptonic force.  Differing from the framework adopted in the
previous works, the leptonic gauge interaction $L_{\mu} - L_{\tau}$ in
our scenario does not discriminate between the charged lepton and the
corresponding neutrino.  We take an advantage of the interaction in the
charged lepton sector in order to account for the inconsistency in the
$g_{\mu} -2$.  In short, we examine if this leptonic force
simultaneously explains the two phenomena in the lepton sector: the gap
in the cosmic neutrino spectrum and the long-standing inconsistency in
the $g_{\mu} -2$.  
It has been pointed out~\cite{Altmannshofer:2014pba} that one of 
the attractive scenarios to solve the $g_{\mu} -2$ problem
is a new muonic force mediated by a field with a mass of
$\mathcal{O}(1)$ MeV, which is, by accident, within the mass range 
of the mediation field of the neutrino
secret interaction that can attenuate the cosmic neutrinos with energy
around $\mathcal{O}(1)$
PeV~\cite{Ng:2014pca,Ioka:2014kca,Ibe:2014pja,Blum:2014ewa}.  We will
demonstrate that the strength of the leptonic force, which can explain
the observed value of the $g_{\mu} -2$, reproduces the gap in the {\sf
IceCube} spectrum.  
It is also interesting to point out that the model
parameters in our scenario are manifestly related to the distance to the
source of the high-energy cosmic neutrinos.  We will briefly discuss
this point later.

\vspace{3mm}
{\it Model --}
We consider the following gauge interactions:
\begin{eqnarray}
{\cal L}_{Z^\prime}=
g_{Z^\prime}^{} Q_{\alpha\beta}(
  \overline{\nu_{\alpha}}\gamma^\rho P_L \nu_{\beta} 
+\overline{\ell_\alpha}\gamma^\rho \ell_\beta 
) Z^\prime_\rho  
~,
\label{eq:Lz}
\end{eqnarray}
where $Z^\prime$ is the new gauge boson with the gauge coupling
$g_{Z^\prime}^{}$, 
$\alpha,\beta=e,\mu,\tau$, 
and $Q_{\alpha\beta}={\rm diag}(0,1,-1)$ 
represents the charge matrix of $L_\mu - L_\tau$.
After $L_\mu - L_\tau$ is spontaneously broken, $Z^\prime$ acquires 
a mass, $m_{Z^\prime}$.
In order to keep generality, however, we do not go into the details 
of the symmetry breaking and simply treat $m_{Z^\prime}$ 
as a model parameter. 
Also, the kinetic mixing with the SM $U(1)_Y$ is set to zero.
The first term of Eq. (\ref{eq:Lz}) is the source of the secret 
neutrino interaction.
In the $L_\mu - L_\tau$ model, as discussed in the next section, 
a mean free path (MFP) of the cosmic neutrino is calculated to 
be $>{\cal O}(1)$ ${\rm Mpc}$, which is many orders of magnitude 
larger than the coherence length.
Travelling such a long distance, neutrino flavor eigenstates
are expected to lose their coherence, and thus the scattering 
process can be described in terms of mass eigenstates with 
the Lagrangian
\begin{eqnarray}
{\cal L}_{Z^\prime \nu \nu}=
g_{ij}^\prime~
\overline{\nu_{i}}\gamma^\rho P_L \nu_{j} Z^\prime_\rho  
~,
\label{eq:nSI}
\end{eqnarray}
where $g_{ij}^\prime=g_{Z^\prime} (V^\dag Q V)$ with $i,j=1\cdots 3$,
and $V$ is the lepton mixing matrix.
In order to realize the gap in the cosmic neutrino spectrum, we utilize a resonant interaction and take a Breit-Wigner form.
Then the scattering cross section of a $\nu_i\overline{\nu_j} \rightarrow \nu \overline{\nu}$ process is obtained as
\begin{eqnarray}
\sigma_{ij} = \frac{1}{6\pi} |g^\prime_{ij}|^2  g_{Z^\prime}^2 
\frac{s}{(s-m_{Z^\prime}^2)^2 + m_{Z^\prime}^2 \Gamma_{Z^\prime}^2}
~,
\end{eqnarray}
where $\sqrt{s}$ is the center-of-mass energy 
and $\Gamma_{Z^\prime}=g_{Z^\prime}^2 m_{Z^\prime}/(12\pi)$ 
is the decay width of $Z^\prime$.  
Throughout this study, we use $g^\prime_{ij}$ evaluated with the 
best fit values of the neutrino mixing parameters \cite{Forero:2014bxa}:
\begin{eqnarray}
\frac{|g_{ij}^\prime|}{ g_{Z^\prime}}=
\left( \begin{array}{ccc}
0.054(0.051) & 0.163(0.158) & 0.555(0.556) \\
0.163(0.158) & 0.088(0.082) & 0.806(0.808) \\
0.555(0.556) & 0.806(0.808) & 0.143(0.133) 
\end{array} \right)~~
\label{eq:coupling}
\end{eqnarray}
for the inverted (normal) mass hierarchy, IH (NH).  
For the mass-squared differences, we also use the best 
fit values~\cite{Forero:2014bxa}.
We take into account 
the constraint from cosmology on the sum of 
the neutrino masses, 
$\sum_{i} m_{\nu_i} \lesssim 0.3$ 
eV~\cite{Hinshaw:2012aka, Ade:2013zuv, Mantz:2014paa}.
Note that all the elements of $g_{ij}'$ are not vanishing, 
which means that each mass eigenstate of the cosmic neutrinos 
can be attenuated by all mass eigenstates of the C$\nu$B.
This is one of the distinctive features of our scenario.

In analogy with the previous works~\cite{Ibe:2014pja,Blum:2014ewa}, 
we assume that the ratio of initial fluxes in the flavor basis is 
$\phi_e : \phi_\mu : \phi_\tau=1:2:0$, 
which is converted into that in the mass basis via 
$\phi_i \equiv \sum_{\beta} |V_{\beta i}|^2 \phi_\beta$.
In view of $\theta_{13}\simeq 0$ and $\theta_{23}\simeq \pi/4$, 
it is reasonable to approximate $\phi_1 : \phi_2 : \phi_3=1:1:1$ 
for simplicity, and indeed we assume this ratio throughout 
this study.
Note that our results are not largely affected by the changes 
of the initial flux ratio, since all mass eigenstates of the cosmic 
neutrinos can be attenuated by one C$\nu$B state.

\begin{figure}[t]
\begin{center}
\includegraphics[width=6.5cm,clip]{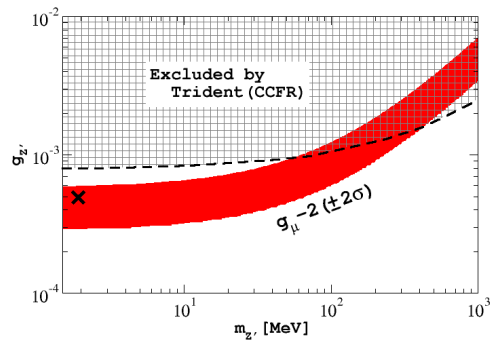} 
\caption{%
The shaded (red) band is 
the $\pm 2 \sigma$ parameter space 
for the $g_\mu-2$~\cite{Hagiwara:2011af}.  
The hatched (gray) region is excluded by the constraint
from the neutrino trident production process
at $95\%$ C.L.~ \cite{Altmannshofer:2014pba}. 
The symbol ``$\times$'' indicates the set of parameters 
used in Figs.~\ref{fig:MFP}, \ref{fig:MFPz}, 
\ref{fig:variation}, and \ref{fig:flux} as reference.
}
\label{fig:mz-gp}
\end{center}
\end{figure}
The introduction of the $L_\mu - L_\tau$ symmetry brings 
not only the secret neutrino interaction but also the new 
interaction among the charged leptons.  
This gives us a chance to solve the inconsistency in 
the $g_\mu-2$~\cite{Baek:2001kca}.
In Fig.~\ref{fig:mz-gp}, we show the parameter region favored 
by the observations of the $g_{\mu} -2$ within $2\sigma$ 
with the shaded (red) band \cite{Hagiwara:2011af}.
The region excluded by 
the neutrino trident production process \cite{Altmannshofer:2014pba} 
from the CCFR experiment \cite{Mishra:1991bv}
is also indicated with the hatched (grey) region.
We will demonstrate that 
the gap is successfully reproduced
with the parameters in the shaded (red) region.
In the next section, 
we will calculate the flux of the cosmic neutrinos with 
$(g_{Z^\prime},m_{Z^\prime}) = (5\times 10^{-4},1.9\ {\rm MeV})$ 
for the IH case, which is  represented by ``$\times$''  
in Fig.~\ref{fig:mz-gp}.

\vspace{3mm}
{\it Result -- }
We consider that the cosmic neutrinos, $\nu_i$, 
are attenuated by the interaction Eq.~\eqref{eq:nSI} with 
the C$\nu$B, $\bar{\nu}_{j}$. 
As reference, in what follows, 
we will use $z=0.2$ and  $m_{\nu_{3}}= 3 \times 10^{-3}$ eV for IH.
As for the NH case, several comments are given at the end of this section. 
The MFP $\lambda_{i}$ of the cosmic neutrino $\nu_{i}$ 
with energy $E_{\nu_i}$ is described as
\begin{equation}
 \lambda_i (E_{\nu_i}, z) = 
\left[
\sum_{j=1}^{3} \int
\frac{d^3 \bm{p}}{(2 \pi)^3} f_j(|\bm{p}|,z) 
\sigma_{ij} (\bm{p},E_{\nu_i}^s)
\right]^{-1},
\label{eq:MFP}
\end{equation}
where $z$ is the parameter of redshift, 
$\boldsymbol{p}$ is the momentum of the C$\nu$B, and 
$f_j(|\bm{p}|,z) =
(e^{|\bm{p}|/(T_{\nu 0} (1 + z))}+ 1)^{-1}$ is the distribution function
with the C$\nu$B temperature $T_{\nu 0} \sim 1.95$ K at present.
Note that $E_{\nu_i}^{s}$ is the energy of a cosmic neutrino 
$\nu_{i}$, which is measured at the position $z$ where the $\nu_{i}$ 
is scattered, and $E_{\nu_{i}}$ is the energy measured 
at {\sf IceCube}~\cite{Yoshida:2012gf,Yoshida:2014uka}.
They are related as 
$E_{\nu_{i}}^{s} = (1+z)E_{\nu_{i}}$.
The survival rate $R_{i}$ of the cosmic neutrino $\nu_{i}$ 
travelling from the source at $z$ to us $(z=0)$ is evaluated by
\begin{equation}
R_i=
\exp
\left[
- \int_0^z \frac{1}{\lambda_i (E_{\nu_i},z')} \frac{d L}{d z'} dz'
\right],
\label{eq:flux}
\end{equation}
where ${d L}/{d z} = c/(H_0\sqrt{\Omega_m(1+z)^3 + \Omega_\Lambda})$
with the present values of the cosmological parameters;
the matter energy density $\Omega_m = 0.315$, 
the dark energy density $\Omega_\Lambda = 0.685$, 
and the Hubble constant $H_0 = 100h$ km/s/Mpc with $h =
0.673$~\cite{PDG:2014}.
Here we assume that the source of cosmic neutrinos 
is located at a particular $z$. 
We expect that the effect caused by the inclusion of realistic 
(widespread) distribution of neutrino sources 
is limited in our case since $z$ is chosen to be small.
\begin{figure}[t]
\begin{center}
\includegraphics[clip,width=0.8\linewidth]{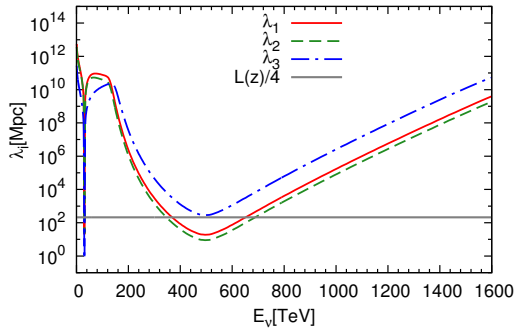} 
\end{center}
\caption{
The MFPs of the cosmic neutrinos for the IH case. 
The solid (red), dashed (green) and dash-dotted (blue) curves 
correspond to the MFPs of the cosmic neutrinos, 
$\nu_1$, $\nu_2$ and $\nu_3$, respectively.
The horizontal (gray) line stands for $L/4$ as reference.
}
\label{fig:MFP}
\end{figure}
We present an example of the MFPs of each mass eigenstate
of the cosmic neutrinos with a fixed value $0.2$ of $z$
in Fig.~\ref{fig:MFP} for the IH case.
When the resonance condition $s\simeq m_{Z'}^2$ in $\sigma_{ij}$ 
is satisfied, the MFP takes its minimum value. 
With the energy $E_{\nu_{i}}^{\text{res}}$ at which
the resonance takes place,
the condition for the cosmic neutrino $\nu_{i}$ 
is described as
\begin{equation}
 m_{Z^\prime}^2 \simeq 
2 E_{\nu_i}^{\text{res}} (1+z) 
\left[
\sqrt{|\bm{p}|^2 + m_{\nu_j}^2} - |\bm{p}| \cos \theta
\right],
\label{eq:SCOME}
\end{equation}
where $\theta$ is the angle between the momentum of the cosmic neutrino
and that of the C$\nu$B.
Applying the parameters adopted in
Fig.~\ref{fig:MFP} to Eq.~\eqref{eq:SCOME},
we find the following two resonant energies
\begin{align}
 E_{\nu_i}^{\text{res}} 
 = 
 \begin{cases}
  \frac{1}{1+z}\frac{m_{Z'}^2}{2m_{\nu_{1(2)}}}
 \simeq 30\ 
  {\rm TeV}, 
 \\
 \frac{1}{1+z}\frac{m_{Z'}^2}{2m_{\nu_3}} 
 \simeq 500\ {\rm TeV}~.
 \end{cases}
\end{align}
Indeed, one can see the two resonance structures 
in Fig.~\ref{fig:MFP}.
Note that the dip around $500$ TeV on each MFP $\lambda_{i}$ 
is created by the scattering with the lightest 
C$\nu$B state $\bar{\nu}_{3}$,
and a narrow dip around $30$ TeV consists of 
the contributions from two heavy states 
$\bar{\nu}_{1}$ and $\bar{\nu}_{2}$.
The resonant condition Eq.~\eqref{eq:SCOME}
would help us reproduce the {\sf IceCube} gap 
at the suitable energy range in the calculation 
of the total flux which will be shown below.

\begin{figure}[t]
\begin{center}
\includegraphics[clip,width=0.8\linewidth]{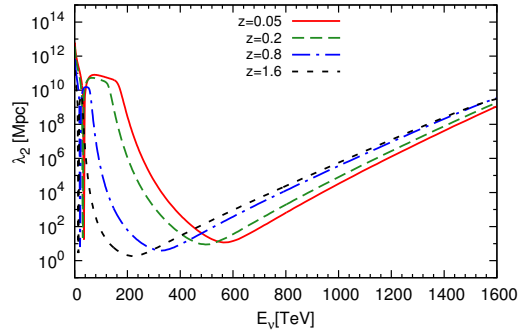} 
\end{center}
\caption{%
The MFPs of the cosmic neutrino $\nu_2$ for the IH case 
with the various $z$.
}
\label{fig:MFPz}
\end{figure}
\begin{figure}[t]
\begin{center}
 \includegraphics[clip,width=0.8\linewidth]{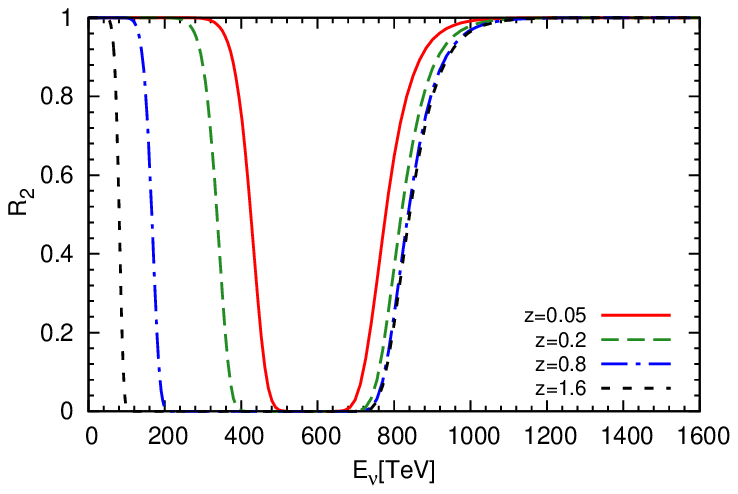} 
 \includegraphics[clip,width=0.8\linewidth]{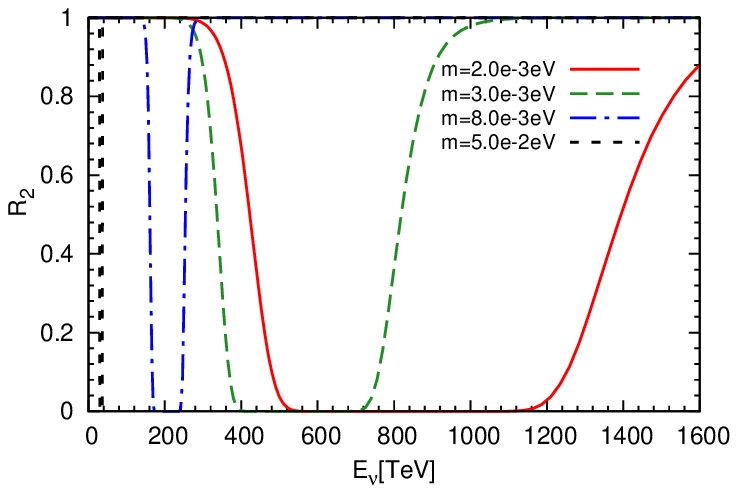} 
\end{center}
\caption{ 
The survival rates of the cosmic neutrino $\nu_2$ for the IH case 
with the various $z$ (upper panel) and $m_\nu$ (lower panel).
Only the scattering between the cosmic neutrino $\nu_2$ and the 
C$\nu$B $\bar{\nu}_3$ is considered.
}
\label{fig:variation}
\end{figure}

Let us explain four important points to
understand the features of the MFPs shown in Fig.~\ref{fig:MFP}.
Firstly, each mass eigenstate of the cosmic neutrinos 
can be attenuated by all mass eigenstates of the C$\nu$B 
through the couplings given in Eq.~\eqref{eq:coupling}.
Secondly, the difference in the depth of the MFPs 
at each resonance energy stems from 
the difference of the couplings $g_{ij}'$ 
for each combination of $i$ and $j$.
For example, at the resonance around $500$ TeV 
in Fig.~\ref{fig:MFP}, 
all mass eigenstates $\nu_{1,2,3}$ of the cosmic neutrinos 
are attenuated by the lightest C$\nu$B, i.e., $\bar{\nu}_3$.
Because of $|g_{33}^\prime| < |g_{23}^\prime|, |g_{13}^\prime|$ 
(see Eq.~\eqref{eq:coupling}), 
the MFP $\lambda_{3}$ for $\nu_3$ becomes longer 
than the others $\lambda_{1,2}$.
Although the strength of the interactions between $\nu_{3}$ and
$\bar{\nu}_{3}$ is relatively small, it is still large enough to 
scatter off $\nu_{3}$ around the resonance.
Thirdly, the thermal distribution effect of 
the C$\nu$B becomes more important for the C$\nu$B  
with a smaller mass, which is apparent from Eq.~\eqref{eq:SCOME}. 
The effect makes the energy range at which
the resonance condition is satisfied broader.
This feature can be clearly read off from Fig.~\ref{fig:MFP}: 
the resonance around $500$ TeV, 
which is associated with the lightest C$\nu$B state, 
is broader than that around $30$ TeV, 
which is related to the heavier C$\nu$B states.
Finally, we pay attention to the fact that
the resonance energy measured at {\sf IceCube} 
depends on $z$, as described at 
Eq.~\eqref{eq:SCOME}.
The observed gap in the spectrum results from 
superposition of the resonant effect in the MFP 
with different $z$ along the path of the cosmic neutrino
from the source to {\sf IceCube}.
To investigate the resonance energy range from another point of view, 
we draw the MFPs $\lambda_{2}$ for $\nu_{2}$ 
with various values of $z$ in Fig.~\ref{fig:MFPz}, 
where the other parameters are fixed. 
As is expected from the $z$-dependence of the distribution function 
$f_{j}$,  a larger value of $z$ makes the resonance region broader.
The position of the resonant energy varies 
with the change of $z$ along the path of the cosmic neutrino.
This behaviour can be understood from the 
redshift of the resonant energy, cf. Eq.~\eqref{eq:SCOME}.
From these two effects brought by $z$,
we expect that a choice of a larger value of $z$
makes the gap width broader in the calculation 
of the total flux.
Since the width of the gap is determined mainly by 
the smallest neutrino mass and the distance $z$ 
to the neutrino source, there is a strong correlation between 
them.
For example, when $z$ is taken to be small, 
the lightest neutrino mass must also be small.
In terms of the survival rate $R$, we can also confirm this correlation 
in Fig.~\ref{fig:variation}, where only the scattering between the 
cosmic neutrino $\nu_2$ and the C$\nu$B $\bar{\nu}_3$ is considered.
In the upper (lower) panel, $R_2$ is calculated with various 
values of $z$ ($m_{\nu}$), while fixing the other parameters 
at their reference values.
These plots tell us that when we know the distance ($z$) to 
the neutrino source, 
we can predict the neutrino mass associated with the gap.
By scanning the values of $z$ and $m_{\nu_i}$, we find that 
the lightest neutrino mass should be larger 
than ${\cal O}(10^{-3})$ eV in a small $z$ region.

\begin{figure}[t]
\begin{center}
\includegraphics[clip,width=0.8\linewidth]{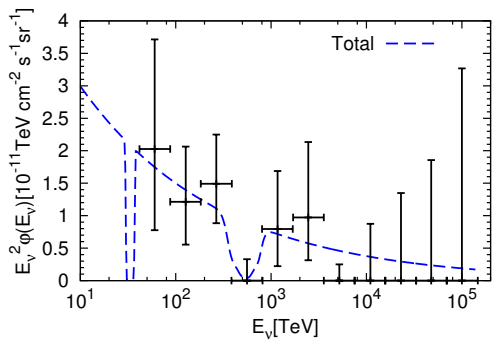}
\end{center}
\caption{%
The total flux of the cosmic neutrinos $(\nu + \bar{\nu})$ for the IH case.
}
\label{fig:flux}
\end{figure}

In Fig.~\ref{fig:flux},
we, finally, calculate the total flux
$\varphi(E_{\nu})$ of
the cosmic neutrinos $(\nu + \bar{\nu})$ 
with the set of the parameters, 
which solves the $g_\mu-2$ problem. 
Although we have an additional resonance caused by the interaction 
with the other mass eigenstates of C$\nu$B at the low energy region, 
it may be difficult to observe such a narrow resonance structure
at the low energy region in the present {\sf IceCube} data, which is
easily smeared by atmospheric neutrino events.
We have also checked if a realistic gap is obtained for the NH case as well
by using similar values of the parameters, but found that it is difficult.
This is because a gap caused by the lightest C$\nu$B $\bar{\nu}_1$ 
is always accompanied by that caused by $\bar{\nu}_2$, and the latter 
attenuates the flux around $400$ TeV.
It may be interesting to focus on large neutrino mass regions, so that 
the lighter (or all) neutrino masses are degenerated, and thus the two 
(or three) gaps are merged into one gap between $400$ TeV - $1$ PeV.
We will study this possibility elsewhere.

\vspace{3mm}
{\it Summary -- }
We have discussed the possibility whether the gap in the cosmic neutrino flux 
within the energy range $400$ TeV - $1$ PeV reported by the {\sf IceCube} 
experiment can be explained in the gauged $L_\mu - L_\tau$ model.
We have shown that the MFPs of the cosmic neutrinos can be 
reduced enough by the resonant scattering 
with the lightest C$\nu$B state for IH.
We have also shown that the MFP has 
a dip with an appropriate width 
for the lightest neutrino mass 
with a value around several $10^{-3}$ eV. 
This is because the thermal distribution of C$\nu$B makes 
the resonant energy range satisfying 
Eq.~\eqref{eq:SCOME} broader.
The dip in the flux becomes broader as the redshift is higher, 
also because of the effect of superposition of the MFPs with different $z$
(cf.~Fig.~\ref{fig:MFPz}).
Once a value of $m_{\nu}$ is fixed, 
the redshift is determined so as to explain the observed gap 
(cf. Fig.~\ref{fig:variation}).
In Fig.~\ref{fig:flux}, we have shown that observed gap 
in the cosmic neutrino spectrum is obtained 
for the lightest neutrino mass $3 \times 10^{-3}$ eV and 
the $Z'$ boson mass $1.9$ MeV for the IH case. 
The gauge coupling constant is taken as 
$5 \times 10^{-4}$, which can settle the $g_\mu - 2$ problem. 
Importantly, in this example, the redshift is determined 
as $0.2$, which corresponds to about  $0.845$ Gpc to neutrino sources. 
In contrast, for the NH case, it is difficult to reproduce a realistic gap 
with a similar set of the parameters.

Before closing the summary, three comments are in order. 
(1) With the neutrino masses and the mixing parameters applied in our 
analysis, the effective neutrino mass of neutrinoless double decay 
processes, $\langle m_{ee} \rangle$, is between 
$4.81 \times 10^{-2}$ and $1.67 \times 10^{-2}$ eV in the IH case, 
which will be examined by the {\sf KamLAND-Zen} 
experiment \cite{TheKamLAND-Zen:2014lma}. 
(2)
Also the sum of the neutrino masses, $\sum m_{\nu}$, is 
$0.102$ eV, which will be explored in future astrophysical 
observations \cite{Abazajian:2013oma}.
(3)
In this study, the effect caused by the neutrinos after 
the scattering was not taken into account.
Inclusion of the secondary neutrinos may explain 
a small bump at the lower energy bin next to the gap.
We also did not consider the distribution of sources of cosmic
neutrinos. The impact of the source distribution on 
our results may be limited, because the distance to the source 
was taken to be small.
We are preparing for a detailed study with a large parameter scan, 
full consideration of the source distribution and 
the secondary neutrino effect.

\vspace{3mm}
{\it Acknowledgments --} 
We would like to thank Professor Shigeru Yoshida for 
insightful comments and careful reading of the manuscript.
The research of F.K. is financially supported by the Sasakawa 
Scientific Research Grant from The Japan Science Society.
This work is supported in part by JSPS KAKENHI
No.~26105503 (T.O.), No.~24340044, No.~25105009 (J.S.) 
and No.~23740190 (T.S.).


\newpage
\bibliography{./IceCubeGap}
\bibliographystyle{apsrev}

\end{document}